\documentclass[aps,preprint]{revtex4}%
\usepackage{amsfonts}
\usepackage{amsmath}
\usepackage{amssymb}
\usepackage{graphicx}%
\providecommand{\U}[1]{\protect\rule{.1in}{.1in}}

\begin{document}
\preprint{ }
\title{Use of a physical metric for OPERA experiment}
\author{Yukio Tomozawa}
\email{tomozawa@umich.edu}
\affiliation{Michigan Center for Theoretical Physics and Randall Laboratory of Physics,
University of Michigan, Ann Arbor, MI. 48109-1120, USA}
\author{}
\date{\today}

\begin{abstract}
A physical metric is introduced as one that directly gives experimental data
without further coordinate transformation. It will be shown that the geodesic
equation for the Schwartzschild metric does not give the correct expression
for the Shapiro time delay experiment. In the physical metric the speed of
light on the surface of the earth is exactly the same as that in vacuum to
first order in gravity, and this should eliminate the inconsistency of the
speed of the neutrino exceeding that of light, if enough care is exerted in
the discussion.

\end{abstract}

\maketitle

\section{Introduction}

It is well known that the coordinates of the Schwartzschild metric do not
correspond to observable physical coordinates\cite{mtw},\cite{sw}. The author
has introduced a coordinate transformation for the Schwartzschild metric, so
that the geodesic equation for the time delay of light propagation in a
gravitational field gives the correct result for Shapiro's
experiment\cite{shapiro 2}. We call such a metric a physical metric. We start
by constructing such a metric.

\section{Asymptotic form for the physical metric}

The physical metric is expressed as
\begin{equation}
ds^{2}=e^{\nu(r)}dt^{2}-e^{\lambda(r)}dr^{2}-e^{\mu(r)}r^{2}(d\theta^{2}%
+\sin^{2}\theta d\phi^{2}),
\end{equation}
for a spherically symmetric and static mass point $M$. From the fact that the
transformation, $r^{\prime}=re^{\mu(r)/2}$, leads to the Schwartzschild
metric, one can deduce the expression for the metric,%
\begin{equation}
e^{\nu(r)}=1-(r_{s}/r)e^{-\mu(r)/2}, \label{metric1}%
\end{equation}%
\begin{equation}
e^{\lambda(r)}=(\frac{d}{dr}(re^{\mu(r)/2}))^{2}/(1-(r_{s}/r)e^{-\mu(r)/2}),
\label{metric2}%
\end{equation}
where $r_{s}=2GM/c^{2}$ is the Schwartzschild radius. An asymptotic expansion
for the metric functions can be obtained from Eq. (\ref{metric1}) and Eq.
(\ref{metric2}), yielding%
\begin{equation}
e^{\nu(r)}=\sum_{n=0}^{\infty}a_{n}(r_{s}/r)^{n},\text{ }e^{\lambda(r)}%
=\sum_{n=0}^{\infty}b_{n}(r_{s}/r)^{n},\text{ }and\text{\ \ }e^{\mu(r)}%
=\sum_{n=0}^{\infty}c_{n}(r_{s}/r)^{n}, \label{eqasympt}%
\end{equation}
where%
\begin{align}
a_{0}  &  =b_{0}=c_{0}=1,\label{eq3}\\
-a_{1}  &  =b_{1}=1\text{ \ \ }and\label{eq4}\\
a_{2}  &  =c_{1}/2,\text{ \ }b_{2}=1-c_{1}/2+c_{1}^{2}/4-c_{2},\text{ \ }etc.
\label{eq5}%
\end{align}
It is obvious that $a_{n+1\text{ }}$and $b_{n}$ can be expressed as functions
of $c_{n}$, $c_{n-1}$ $\ldots$, $c_{1}$.

\section{Geodesic equations and time delay}

The geodesic equations can be obtained from variations of the line integral
over an invariant parameter $\tau$,$\ \int(\frac{ds}{d\tau})^{2}d\tau$, and
their integrals are given by%

\begin{equation}
\frac{dt}{d\tau}=e^{-\nu(r)}, \label{eq1}%
\end{equation}

\begin{equation}
\frac{d\phi}{d\tau}=J_{\phi}e^{-\mu(r)}/(r\sin\theta)^{2},
\end{equation}

\begin{equation}
(\frac{d\theta}{d\tau})^{2}=(J_{\theta}^{\text{ }2}-J_{\phi}^{\text{ }2}%
/\sin^{2}\theta)e^{-2\mu(r)}/r^{4}.
\end{equation}
Restricting the plane of motion to $\frac{d\theta}{d\tau}=0,$ $\theta=\pi/2,$
the radial part of the geodesic integral is given by%

\begin{equation}
(\frac{dr}{d\tau})^{2}=e^{-\lambda(r)}(e^{-\nu(r)}-J^{\text{ }2}e^{-\mu
(r)}/r^{2}-E) \label{eq2}%
\end{equation}
where $J_{\phi}$, $J_{\theta}$ and $E$ are constants of integration and%

\begin{equation}
J^{\text{ }2}=J_{\phi}^{\text{ }2}=J_{\theta}^{\text{ }2}.
\end{equation}
The constant $E$ is 0 for light propagation.

\bigskip From Eq. (\ref{eq1}) and Eq. (\ref{eq2}) with Eqs. (\ref{eq3}) and
(\ref{eq4}), it follows that%
\begin{align}
\frac{dt}{dr}  &  =\pm\text{ }e^{-\nu(r)}/\sqrt{e^{-\nu(r)-\lambda
(r)}-J^{\text{ }2}e^{-\mu(r)-\lambda(r)}/r^{2}}\\
&  =\pm\text{ }\frac{rr}{\sqrt{r^{2}-r_{0}^{\text{ }2}}}\text{ }%
(1+\frac{(b_{1}-a_{1})\text{ }r_{s}}{2r}+\frac{(c_{1}-a_{1})\text{ }%
r_{0}\text{ }r_{s}}{2\text{ }r\text{ }(r+r_{0})}+\cdots)
\end{align}
for light propagation, where $r_{0}$ is the impact parameter, and%
\begin{equation}
J^{2}=r_{0}^{2}e^{-\nu(r_{0})+\mu(r_{0})}.
\end{equation}
Integrating from $r_{0}$ to $r$, one gets the time delay expression for light
propagation,%
\begin{equation}
\bigtriangleup t=\text{ }r_{s}\text{ }(\ln(\frac{r+\sqrt{r^{\text{ }2}%
-r_{0}^{\text{ \ }2}}}{r_{0}})+\frac{(c_{1}+1)}{2}\sqrt{\frac{r-r_{0}}%
{r+r_{0}}})+\cdots. \label{eq6}%
\end{equation}
\ For the Schwartzschild metric, $c_{1}=0$, the time delay is given by
\begin{equation}
\bigtriangleup t=\text{ }r_{s}\text{ }(\ln(\frac{r+\sqrt{r^{\text{ }2}%
-r_{0}^{\text{ \ }2}}}{r_{0}})+\frac{1}{2}\sqrt{\frac{r-r_{0}}{r+r_{0}}%
})+\cdots,
\end{equation}
which agrees with the calculation of Weinberg\cite{sw2}

\bigskip Since Shapiro's observation\cite{shapiro 2} fits the formula%

\begin{equation}
\bigtriangleup t=\text{ }r_{s}\text{ }\ln(\frac{r+\sqrt{r^{\text{ }2}%
-r_{0}^{\text{ \ }2}}}{r_{0}})+\cdots\label{eq7}%
\end{equation}
to high accurracy (1 in 1000), we conclude that
\begin{equation}
c_{1}=-1. \label{eq8}%
\end{equation}

\bigskip We note that the parameter values%
\begin{equation}
a_{1}=-1,\text{ }and\text{ }b_{1}=1 \label{eq8.5}%
\end{equation}
are coordinate independent and determined from the solution of the Einstein
equation and the physical boundary condition. Thus we conclude that Eq.
(\ref{eq8}), along with Eq. (\ref{eq8.5}), is the condition for the physical
metric. It is important to notice that the speed of light on the surface of a
sphere is identical to that in vacuo, $c$, in the physical metric, while it is
less than $c$ in the Schwartzschild metric,%
\begin{equation}
c_{S}/c=(1-r_{s}/r)^{1/2},
\end{equation}
where
\begin{equation}
r_{s}=\frac{2GM_{E}}{c^{2}}=0.886\text{ }cm
\end{equation}
and%
\begin{equation}
r_{s}/r=1.39\text{ }10^{-9}%
\end{equation}
on the surface of the Earth.

\section{The speed of neutrinos in the OPERA experiment}

The speed of a neutrino\cite{opera} of 17 GeV with the assumed mass of the
neutrino, $m_{\nu}c^{2}=0.1$ $eV$, can be calculated from%
\begin{equation}
E_{\nu}=\frac{m_{\nu}c^{2}}{\sqrt{1-(v/c)^{2}}}=17GeV
\end{equation}
resulting in%
\begin{equation}
\frac{v}{c}=1-1.73\text{ }10^{-23}.
\end{equation}
Obviously%
\begin{equation}
c>v>c_{S}=c\text{ }(1-0.695\text{ }10^{-9})
\end{equation}
It is clear that if the physical metric were to be used from the beginning,
the problem of the speed of the neutrino exceeding that of the light would not
occur. It exceeds the speed of light in the Schwartzschild metric, $c_{S}$,
but that conflicts with the time delay experiment of Shapiro. Whenever the
gravitational time delay formula is used in GPS operation, you might be using
the Schwartzschild metric unless extreme care is utilized. One way to remedy
the deficiency is to reduce distances on the surface of the earth by a factor
of $(1-0.695\text{ }10^{-9})$, the same factor as that of the gravitational
time delay on the Earth, so that the speed of light becomes that in vacuum,
$c$.

Shapiro's time delay experiment has been successfully applied to a neutron
star binary\cite{nsb}. This implies that the formula in Eq. (\ref{eq7}) is
valid to higher order in gravity. Then, the physical metric to higher order
may be neccessary and important. For such a work, see the reference of the
present author\cite{higher}.

\begin{acknowledgments}
It is a pleasure to thank David N. Williams for reading the manuscript.
\end{acknowledgments}


\begin{thebibliography}{9}                                                                                                %


\bibitem {mtw}C. W. Misner, K. S. Thorne and J. A. Wheeler, Gravitation,
(Freeman, San Francisco, 1973)

\bibitem {sw}S. Weinberg, Gravitation and Cosmology, (Wiley and Sons, New
York, 1972)

\bibitem {shapiro 2}R. D. Resenberg and I. I. Shapiro, Apj. 234, L219 (1979)

\bibitem {sw2}S. Weinberg, Gravitation and Cosmology, (Wiley and Sons, New
York, 1972) p202

\bibitem {opera}T. Adam et.al., Measurement of the neutrino velocity with the
OPERA detector in the CNGS beam, arXiv:hep-ex/1109.4897 v.2 (2011)

\bibitem {nsb}P. D. Demerest et.al., Nature 467, 1081 (2010)

\bibitem {higher}Y. Tomozawa, Coordinate Independence and a Physical Metric in
Compact Form, arXiv:gr-qc/0405071 (2004)
\end{thebibliography}
\end{document}